\documentclass{revtex4}

\usepackage{amsthm,amssymb,amsmath}				
\usepackage{graphicx,comment}
\usepackage{float}   
\usepackage{xcolor}

\begin{document}

\title{PDM Klein-Gordon particles in Som-Raychaudhuri cosmic string spacetime with
space-like dislocation and a non-uniform magnetic field: spacetime associated degeneracies, vorticity-energy, and charge-energy correlations}
\author{Omar Mustafa}
\email{omar.mustafa@emu.edu.tr}
\affiliation{Department of Physics, Eastern Mediterranean University, G. Magusa, north
Cyprus, Mersin 10 - Turkey.}

\begin{abstract}
\textbf{Abstract:} We explicitly study and report  the effects of the gravitational fields on the spectroscopic structure of Klein-Gordon (KG) particles in Som-Raychaudhuri cosmic string spacetime with space-like dislocation and a non-uniform magnetic field. We extend this study to include the effects of the gravitational fields on  some position-dependent mass (PDM) KG-particles. As effects of the gravitational fields, we report new spacetime associated degeneracies,  vorticity-energy correlations and charge-energy correlations. Such degeneracies and correlations  are observed in the spectroscopic structure of both constant mass KG-particles and PDM KG-particles at hand. 


\end{abstract}

\maketitle

\section{Introduction}

The first cosmological solutions, with rotating matter, of Einstein's field equations of gravitation were suggested by the G\"{o}del spacetime metrics \cite{R1}. A generalization of the G\"{o}del-type spacetime metrics is introduced and shown to provide new solutions for gravity and supergravity theories in different dimensions \cite{R2,R3}. It has been explicitly verified that the G\"{o}del-type spacetime metrics with both flat and non-flat backgrounds always have closed time-like curves or null like curves \cite{R4}.  Rebou\c{c}as and Tiomno \cite{R5} have shown that all spacetime (ST)-homogeneous G\"{o}del-type metrics characterized by the vorticity $\Omega
=\pm \left\vert \Omega \right\vert $ can be transformed (with $\hbar =c=G=1$ units) in cylindrical coordinates \cite{R6,R7,R8,R9,R10} to%
\begin{equation}
ds^{2}=-\left( dt+\alpha \Omega \frac{\sinh ^{2}\left( \tilde{\mu}\,r\right) 
}{\tilde{\mu}^{2}}d\varphi \right) ^{2}+\alpha ^{2}\frac{\sinh ^{2}\left( 2\,%
\tilde{\mu}\,r\right) }{4\,\tilde{\mu}^{2}}d\varphi ^{2}+dr^{2}+dz^{2}.
\label{e1}
\end{equation}
Where $\alpha =1-4\tilde{\mu}$ is the disclination parameter (or the deficit angle of the conical geometry), and $\tilde{\mu}$ is the linear mass density
of the string. It should be noted, moreover, that $0<\alpha <1$ is used in general relativity with cosmic string ($\tilde{\mu}>0$), $\alpha =1$
corresponds to Minkowski flat spacetime, and $\alpha >1$ is used in the geometric theory of defects in condensed matter ($\tilde{\mu}<0$). Yet, at
the limit $\tilde{\mu}\longrightarrow 0$ of the G\"{o}del spacetime metric (\ref{e1}) we obtain the ST-homogeneous Som-Raychaudhuri (SR) solution%
\begin{equation}
ds^{2}=-\left( dt+\alpha \,\Omega \,r^{2}d\varphi \right) ^{2}+\alpha
^{2}\,r^{2}d\varphi ^{2}+dr^{2}+dz^{2}  \label{e2}
\end{equation}
of the Einstein field equations \cite{R11}. It is interesting to know that the original G\"{o}del solutions \cite{R1} are obtained for $\tilde{\mu}%
^{2}=\Omega ^{2}/2$ and $\alpha =1$, the anti-de Sitter solution \cite{R12} for $%
\tilde{\mu}^{2}=\Omega ^{2}$ and $\alpha =1$, and the Minkowski solution for $%
\Omega =0$ and $\alpha =1.$ In the current study, however, we shall also consider a space-like dislocation so that the ST-homogeneous SR-metric with space-like dislocation reads%
\begin{equation}
ds^{2}=-\left( dt+\alpha \,\Omega \,r^{2}d\varphi \right) ^{2}+\alpha
^{2}\,r^{2}d\varphi ^{2}+dr^{2}+\left( dz+\delta d\varphi \right) ^{2},
\label{e3}
\end{equation}%
where $\delta $ denotes space-like dislocation parameter. Hence, the covariant and contravariant metric tensors associated with the
Som-Raychaudhuri spacetime with space-like dislocation are given by%
\begin{equation}
g_{\mu \nu }=\left( 
\begin{tabular}{cccc}
$-1\smallskip $ & $\,0\,$ & $-\alpha \Omega r^{2}$ & $\,0$ \\ 
$0$ & $1\smallskip $ & $0$ & $0\smallskip \smallskip $ \\ 
$-\alpha \Omega r^{2}$ & $\,0$ & $\,\left[ \alpha ^{2}r^{2}\left( 1-\Omega
^{2}r^{2}\right) +\delta ^{2}\right] \,$ & $\delta $ \\ 
$0$ & $0$ & $\delta $ & $1$%
\end{tabular}%
\right) \Longleftrightarrow g^{\mu \nu }=\left( 
\begin{tabular}{cccc}
$\left( \Omega ^{2}r^{2}-1\smallskip \right) $ & $\,0\,$ & $-\frac{\Omega }{%
\alpha }$ & $\frac{\Omega \delta }{\alpha }\smallskip $ \\ 
$0$ & $\,1\smallskip $ & $0$ & $0$ \\ 
$-\frac{\Omega }{\alpha }$ & $\,0$ & $\,\frac{1}{\alpha ^{2}r^{2}}\,$ & $-%
\frac{\delta }{\alpha ^{2}r^{2}}\smallskip $ \\ 
$\frac{\Omega \delta }{\alpha }$ & $0$ & $\,-\frac{\delta }{\alpha ^{2}r^{2}}
$ & $\,\left( 1+\frac{\delta ^{2}}{\alpha ^{2}r^{2}}\right) $%
\end{tabular}%
\right) ,  \label{e4}
\end{equation}
where $\det\left( g_{\mu\nu}\right) =-\alpha ^{2}r^{2}$. The G\"{o}del-type ST-homogeneous SR-spacetime with space-like dislocation metric (\ref{e3}) is the background of the current proposal.

However, in classical mechanics, the introduction of  Mathews and Lakshmanan \cite{R13} position-dependent mass (PDM) oscillator has inspired studies on PDM-systems. Few years later, von Roos \cite{R14} has introduced PDM-Schr\"{o}dinger Hamiltonian in quantum mechanics. Such two prominent PDM-models have subsequently attracted research interest over the last few decades both
in quantum and classical mechanics (e.g., \cite{R15,R16,R17,R18,R19,R20,R21,R22,R23,R24,R25,R26,R27,R28,R29,R30,R31}). Hereby, it has been emphasized that the PDM notion is a metaphoric notion that is manifested by coordinate transformation/deformation \cite{R26,R27,R28,R29,R30,R31}. Such coordinate transformation/deformation would necessarily and effectively change the form of the canonical momentum in classical mechanics and the momentum operator in quantum mechanics. In classical mechanics, for example, the canonical momentum $p=m_{\circ }\dot{u}
$ for a particle of rest mass $m_{\circ }$. Which, under the coordinate transformation%
\begin{equation}
u=\int \sqrt{g\left( x\right) }dx=\sqrt{Q\left( x\right) }%
x\Longleftrightarrow \sqrt{g\left( x\right) }=\sqrt{Q\left( x\right) }\left[
1+\frac{Q^{\prime }\left( x\right) }{2Q\left( x\right) }x\right] ,
\label{e1.1}
\end{equation}%
would imply that $\dot{u}=\sqrt{g\left( x\right) }\dot{x}$ and hence the canonical momentum is now given by $p\left( x\right) =m_{\circ }\dot{u}%
=m_{\circ }\sqrt{g\left( x\right) }\dot{x}$. Where, $g(x)$ and $Q(x)$ are positive-valued scalar multipliers. One would then rewrite the PDM canonical momentum as $p\left( x\right) =M\left( x\right) \dot{x}$, where $%
M\left( x\right) =m_{\circ }\sqrt{g\left( x\right) }$. This what renders the mass to be metaphorically and effectively position-dependent.  In quantum
mechanics on the other hand, the PDM-momentum operator is shown \cite{R26} to be given, in $\hbar =1$ units and in its non-minimal coupling form, by%
\begin{equation}
\hat{p}_{j}\left( \mathbf{r}\right) =-i\left( \partial _{j}-\frac{\partial
_{j}g\left( \mathbf{r}\right) }{4g\left( \mathbf{r}\right) }\right)
\,;\,j=1,2,3,  \label{e1.2}
\end{equation}%
and consequently yields, in its most simplistic one-dimensional form, the von Roos \cite{R14} PDM kinetic energy operator%
\begin{equation}
\hat{T}=\left( \frac{\hat{p}_{x}\left( x\right) }{\sqrt{g\left( x\right) }}%
\right) ^{2}=-g\left( x\right) ^{-1/4}\partial _{x}g\left( x\right)
^{-1/2}\partial _{x}g\left( x\right) ^{-1/4},  \label{e1.3}
\end{equation}%
which is known in the literature as Mustafa and Mazharimousavi's \cite{R19} ordering. In short, moreover, for quantum mechanical particles under effective/metaphoric PDM settings, the textbook momentum operator $\hat{p}%
=-i\partial _{x}$ should be replaced by the PDM momentum operator of (\ref{e1.2}). So should be the case for relativistic PDM quantum particles \cite{R28,R29,R30}. Yet, in the relativistic Dirac and Klein-Gordon (KG) wave equations the assumption that the rest mass energy term $m_{\circ
}\longrightarrow m_{\circ }+m\left( x\right) +S\left( x\right) $ would give
PDM settings \cite{R32,R33,R34,R35,R36,R37} should be abandoned, for it just redefines the Lorentz scalar potential $S\left( x\right) \longrightarrow
m\left( x\right) +S\left( x\right) $ and does not introducing PDM settings as a manifestation of coordinate transformation/deformation. PDM KG-particles moving in the background of  SR-spacetime with space-like dislocation form the complementary part of the current proposal.

The organization of the current proposal is in order. In section 2, we revisit and discuss KG-particles in SR-type cosmic string spacetime with space-like dislocation and a non-uniform magnetic field. We observe that, as a consequence of SR-spacetime structure, KG-oscillators are formed. We found that there are degeneracies associated with spacetime structure (to be called hereinafter spacetime associated degeneracies, STAD), where only part of the energy levels (for a given radial quantum number) merge into the corresponding $S$-state (i.e., states with the magnetic quantum number $\ell =0$ ). Interestingly, moreover, we  report two new correlations:  vorticity-energy correlations and charge-energy correlations. Within such findings we discuss, in section 3, the gravitational field effects on PDM KG-oscillators in SR-type cosmic string spacetime with space-like dislocation and a non-uniform magnetic field. In section 4, we discuss and report on some PDM KG-oscillators in a linear confinement in SR-type cosmic string spacetime with space-like dislocation. Our concluding remarks are given in section 5. Our results are thoroughly discussed in the Appendix section. Therein, we provide an exact solution in terms of confluent hypergeometric series/polynomials and report their close connection and consistency with the biconfluent Heun series/polynomials $H_{B}(1, \beta^{\prime}, \gamma^{\prime},0,y)$ as a new finding.

\section{KG-particles in SR-type cosmic string spacetime with space-like screw dislocation}

The KG-equation for a spin-0 particle in a 4-vector $A_{\mu }$ and a Lorentz scalar potentials $S(r)$ in the G\"{o}del SR-type spacetime with space-like dislocation (\ref{e3}) is given by%
\begin{equation}
\frac{1}{\sqrt{-g}}\tilde{D}_{\mu }\left( \sqrt{-g}g^{\mu \nu }\tilde{D}%
_{\nu }\Psi \right) =\left( m_{\circ }+S(r)\right) ^{2}\Psi ,  \label{e5}
\end{equation}%
where $m_{\circ }$ represents the rest mass energy of the KG-particle, $e$ is its charge, $S\left( r\right) $ is the Lorentz scalar potential, and the gauge-covariant derivative is given by $\tilde{D}_{\mu }=D_{\mu }+\mathcal{F}%
_{\mu }=\partial _{\mu }+\mathcal{F}_{\mu }-ieA_{\mu }$. This would in turn allow us to cast equation (\ref{e5}) as 
\begin{gather}
\left\{ \left( \Omega ^{2}r^{2}-1 \right) \tilde{D}_{t}^{2}-\frac{%
2\Omega }{\alpha }\tilde{D}_{t}\,\tilde{D}_{\varphi }+\frac{2\,\Omega
\,\delta }{\alpha }\tilde{D}_{t}\,\tilde{D}_{z}+\frac{1}{r}\tilde{D}_{r}\,r\,%
\tilde{D}_{r}\right.   \notag \\
\left. +\,\frac{1}{\alpha ^{2}r^{2}}\tilde{D}_{\varphi }^{2}-\frac{2\,\delta 
}{\alpha ^{2}r^{2}}\tilde{D}_{\varphi }\,\tilde{D}_{z}+\,\left( 1+\frac{%
\delta ^{2}}{\alpha ^{2}r^{2}}\right) \tilde{D}_{z}^{2}-\left( m_{\circ
}+S\left( r\right) \right) ^{2}\right\} \Psi =0.  \label{e6}
\end{gather}%
Let us assume that $A_{\mu }=\left( 0,0,A_{\varphi },0\right) $, $\mathcal{F}%
_{\mu }=\left( 0,\mathcal{F}_{r},0,0\right) ;\,\mathcal{F}_{r}=g^{\prime
}\left( r\right) /4g\left( r\right) $, where $g\left( r\right) $ is a dimensionless positive-valued scalar multiplier that introduces PDM settings (\ref{e1.2}). We now define our covariant derivatives as%
\begin{equation}
\tilde{D}_{t}=\partial _{t},\;\tilde{D}_{r}=\partial _{r}+\mathcal{F}_{r},\;%
\tilde{D}_{\varphi }=\partial _{\varphi }-ieA_{\varphi },\;\tilde{D}%
_{z}=\partial _{z},  \label{e7}
\end{equation}%
to imply%
\begin{gather}
\left\{ \left( \Omega ^{2}r^{2}-1\right) \,\partial _{t}^{2}-%
\frac{2\Omega }{\alpha }\left( \partial _{\varphi }-ieA_{\varphi }\right) +%
\frac{2\,\Omega \,\delta }{\alpha }\,\partial _{t}\,\partial _{z}+\frac{1}{r}%
\left( \partial _{r}+\mathcal{F}_{r}\right) \,r\,\left( \partial _{r}-%
\mathcal{F}_{r}\right) +\frac{1}{\alpha ^{2}r^{2}}\,\left( \partial
_{\varphi }-ieA_{\varphi }\right) ^{2}\right.   \notag \\
\left. -\frac{2\,\delta }{\alpha ^{2}r^{2}}\left( \partial _{\varphi
}-ieA_{\varphi }\right) \,\partial _{z}+\left( 1+\frac{\delta ^{2}}{\alpha
^{2}r^{2}}\right) \,\partial _{z}^{2}-\left( m_{\circ }+S\left( r\right)
\right) ^{2}\right\} \Psi =0  \label{e8}
\end{gather}%
Then, we use the wave function%
\begin{equation}
\Psi \left( t,r,\varphi ,z\right) =\exp \left( i\left[ \ell \varphi +k\,z-Et%
\right] \right) \psi \left( r\right) =\exp \left( i\left[ \ell \varphi +kz-Et%
\right] \right) \frac{R\left( r\right) }{\sqrt{r}},  \label{e9}
\end{equation}%
and the assumptions%
\begin{equation}
A_{\varphi }=-\frac{\alpha \,B_{\circ }}{2}\,r^{2}+\frac{\phi _{B}}{2\pi }%
\Longrightarrow eA_{\varphi }=-\frac{\tilde{B}_{\circ }}{2}r^{2}+\tilde{\phi}%
;\;\tilde{\phi}=\frac{e\,\phi _{B}}{2\pi },\;\tilde{B}_{\circ }=\alpha
\,e\,B_{\circ },  \label{e10}
\end{equation}%
where $\phi _{B}$ is an internal quantum magnetic flux that admits constant values and $B_{\circ }$ is the external magnetic field strength \cite{R10,R37,R38}. Under such settings, one obtains%
\begin{equation}
\left\{ \partial _{r}^{2}-\frac{\left( \mathcal{L}^{2}-1/4\right) }{r^{2}}-%
\tilde{\Omega}^{2}r^{2}-M\left( r\right) -2m_{\circ }S\left( r\right)
-S\left( r\right) ^{2}+\lambda \right\} R\left( r\right) =0,  \label{e11}
\end{equation}%
with%
\begin{gather}
M\left( r\right) =-\frac{3}{16}\left( \frac{g^{\prime }\left( r\right) }{%
g\left( r\right) }\right) ^{2}+\frac{1}{4}\frac{g^{\prime \prime }\left(
r\right) }{g\left( r\right) }+\frac{g^{\prime }\left( r\right) }{%
4\,r\,g\left( r\right) },\;\tilde{\Omega}^{2}=\left( \Omega E+\frac{%
eB_{\circ }}{2}\right) ^{2},  \label{e12} \\
\lambda =E^{2}-2\left( \Omega E+\frac{eB_{\circ }}{2}\right) \mathcal{L}%
-\left( k^{2}+m_{\circ }^{2}\right) ;\;\,\tilde{\ell}=\frac{\ell -\tilde{\phi%
}}{\alpha },\;\tilde{\delta}=\frac{\delta }{\alpha },\mathcal{L}^{2}=\left(
\,\tilde{\ell}-\tilde{\delta}\,k\right) ^{2},   \label{e13}
\end{gather}%
and $\ell =0,\pm 1,\pm 2,\cdots $ is the magnetic quantum number. Clearly, we find that KG-oscillators are introduced by the very nature of the G\"{o}del SR-type cosmic string spacetime with space-like dislocation.

At this point, one observes that the second term of $\lambda $ in (\ref{e13}) plays a critical role in removing the degeneracies associated with the irrational magnetic quantum number $\mathcal{L}=\pm
\left\vert \mathcal{L}\right\vert $, even when the magnetic field is off (i.e. when $B_{\circ }=0$). This would suggest that whilst the G\"{o}del SR-type cosmic string spacetime manifestly plays the delicate role of a magnetic field (in lifting the degeneracies associated with $\mathcal{L}=\pm
\left\vert \mathcal{L}\right\vert $), it introduces a different type of degeneracies. Such new spacetime associated degeneracies (STAD) are very much related to $\Omega E=\Omega_{\pm}E_{\pm}$ or  $\Omega E=\Omega_{\mp}E_{\pm}$,  for a given $\mathcal{L}$, where the vorticity $\Omega =\pm \left\vert \Omega \right\vert
=\Omega _{\pm }$ and the relativistic particles energies $E=\pm \left\vert
E\right\vert =E_{\pm }$ \cite{R29,R30}. On the other hand, the central repulsive/attractive core $\left( \mathcal{L}^{2}-1/4\right) /r^{2}$ includes within the effect of the space-like dislocation in the form of a shift in the irrational magnetic quantum number $\mathcal{L}= \tilde{%
\ell}-\tilde{\delta}\,k$.  Therefore, the effect of the gravitational field, introduced by the spacetime structure, on the non-relativistic/relativistic quantum system should be treated with diligence. An illustrative example is in order.

\subsection{KG-oscillators in SR-type cosmic string spacetime with space like dislocation and vorticity-energy correlations}

Let us consider the most simplistic and elementary model of (\ref{e11}) with $B_{\circ }=0=\tilde{\phi}=S\left( r\right) $ and $g\left( r\right) =1$ (i.e., KG-particles with constant mass) to obtain%
\begin{equation}
\left\{ \partial _{r}^{2}-\frac{\left( \mathcal{L}^{2}-1/4\right) }{r^{2}}%
-\,\omega ^{2}r^{2}+\lambda \right\} R\left( r\right) =0;\,\lambda =E^{2}-2\,%
\mathcal{L}\,\Omega \,E-k^{2}-m_{\circ }^{2},\,\omega ^{2}=\Omega ^{2}E^{2}.
\label{e14}
\end{equation}
This effective two-dimensional textbook oscillator would immediately imply that 
\begin{equation}
\lambda =2\,\left\vert \omega \right\vert \left( 2n_{r}+\left\vert \mathcal{L%
}\right\vert +1\right) ,\;R\left( r\right) \sim r^{\left\vert \mathcal{L}%
\right\vert +1/2}\exp \left( -\frac{\left\vert \omega \right\vert r^{2}}{2}%
\right) L_{n_{r}}^{\left\vert \mathcal{L}\right\vert }\left( \left\vert
\omega \right\vert r^{2}\right) ,  \label{e15}
\end{equation}%
and hence%
\begin{equation}
E^{2}-2\,\mathcal{L}\,\Omega \,E-\left( k^{2}+m_{\circ }^{2}\right)
=2\,\left\vert \Omega \,E\right\vert \left( 2n_{r}+\left\vert \mathcal{L}%
\right\vert +1\right) .  \label{e15.2}
\end{equation}%
It is obvious that $\,\left\vert \omega \right\vert =\left\vert \Omega
E\right\vert \geq 0$ is a vital condition so that the radial wave function is finite and square integrable, which in turn introduces two sets of eigenvalues and eigenfunctions corresponding to $\left\vert \Omega
E\right\vert =\Omega _{\pm }E_{\pm }$ and $\left\vert \Omega E\right\vert
=-\Omega _{\mp }E_{\pm }$. That is,%
\begin{equation}
E_{\pm }^{2}-2\,\Omega _{\pm }E_{\pm }\mathcal{N}_{+}-\left(
k_{z}^{2}+m_{\circ }^{2}\right) =0 ;\,\, \left\vert \Omega E\right\vert=\Omega _{\pm }E_{\pm }
\label{e15.4}
\end{equation}%
for $\left\vert \Omega E\right\vert =\Omega _{\pm }E_{\pm }$, and%
\begin{equation}
E_{\pm }^{2}+2\,\Omega _{\mp }E_{\pm }\mathcal{N}_{-}-\left(
k_{z}^{2}+m_{\circ }^{2}\right) =0 ; \,\,  \left\vert \Omega E\right\vert
=-\Omega _{\mp }E_{\pm },
\label{e15.5}
\end{equation}
for $\left\vert \Omega E\right\vert =-\Omega _{\mp }E_{\pm }$, where%
\begin{equation}
\mathcal{N}_{\pm }=2n_{r}+\left\vert \mathcal{L}\right\vert \pm \mathcal{L}%
+1\Rightarrow \mathcal{N}_{\pm }|_{_{\mathcal{L}=\mp|\mathcal{L}|}}=\mathcal{N}_{\mp }|_{_{\mathcal{L}=\pm|\mathcal{L}|}} \,,  \label{e15.3}
\end{equation}

The two sets of energies can be treated in two ways. The first of which is based on the possible combinations of $\left\vert \Omega E\right\vert $ so that%
\begin{equation}
E_{\pm ,1}=\pm \left\vert \Omega \right\vert \mathcal{N}_{+}\pm \sqrt{\Omega
^{2}\,\mathcal{N}_{+}^{2}+m^{2}+k^{2}}\Rightarrow \left\{ 
\begin{tabular}{l}
$E_{+,1}=+\left\vert \Omega \right\vert \mathcal{N}_{+}+\sqrt{\Omega ^{2}\,%
\mathcal{N}_{+}^{2}+m_{\circ }^{2}+k^{2}}\medskip $ \\ 
$E_{-,1}=-\left\vert \Omega \right\vert \mathcal{N}_{+}-\sqrt{\Omega ^{2}\,%
\mathcal{N}_{+}^{2}+m_{\circ }^{2}+k^{2}}$%
\end{tabular}%
\right. ,  \label{e16}
\end{equation}
for $|\omega| =+\Omega _{\pm }E_{\pm } $, and%
\begin{equation}
E_{\pm ,2}=\pm \left\vert \Omega \right\vert \mathcal{N}_{-}\pm \sqrt{\Omega
^{2}\,\mathcal{N}_{-}^{2}+m^{2}+k^{2}}\Rightarrow \left\{ 
\begin{tabular}{l}
$E_{+,2}=+\left\vert \Omega \right\vert \mathcal{N}_{-}+\sqrt{\Omega ^{2}\,%
\mathcal{N}_{-}^{2}+m_{\circ }^{2}+k^{2}\medskip}$ \\ 
$E_{-,2}=-\left\vert \Omega \right\vert \mathcal{N}_{-}-\sqrt{\Omega ^{2}\,%
\mathcal{N}_{-}^{2}+m_{\circ }^{2}+k^{2}}$%
\end{tabular}%
\right. ,  \label{e17}
\end{equation}%
\begin{figure}[!ht!]  
\centering
\includegraphics[width=0.3\textwidth]{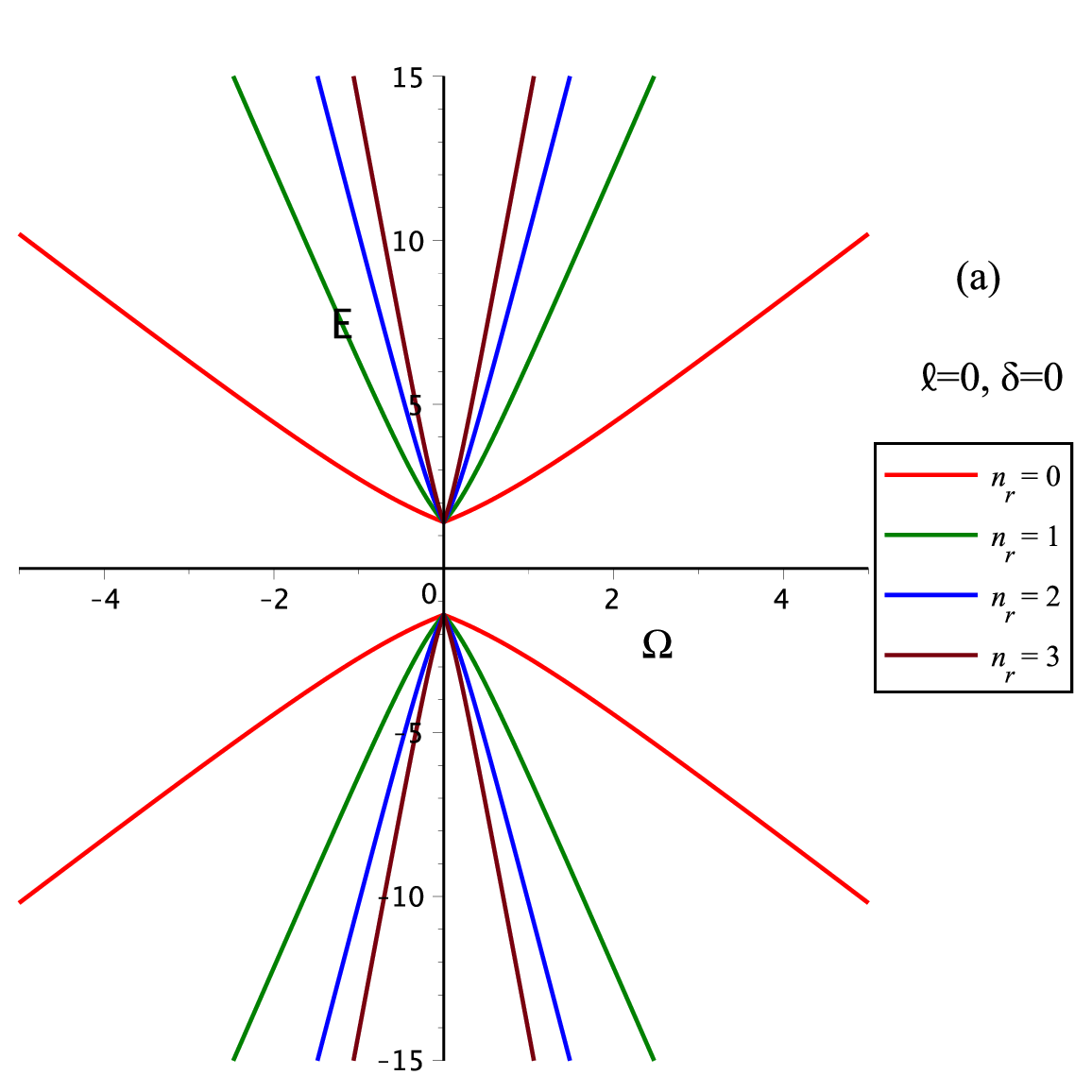}
\includegraphics[width=0.3\textwidth]{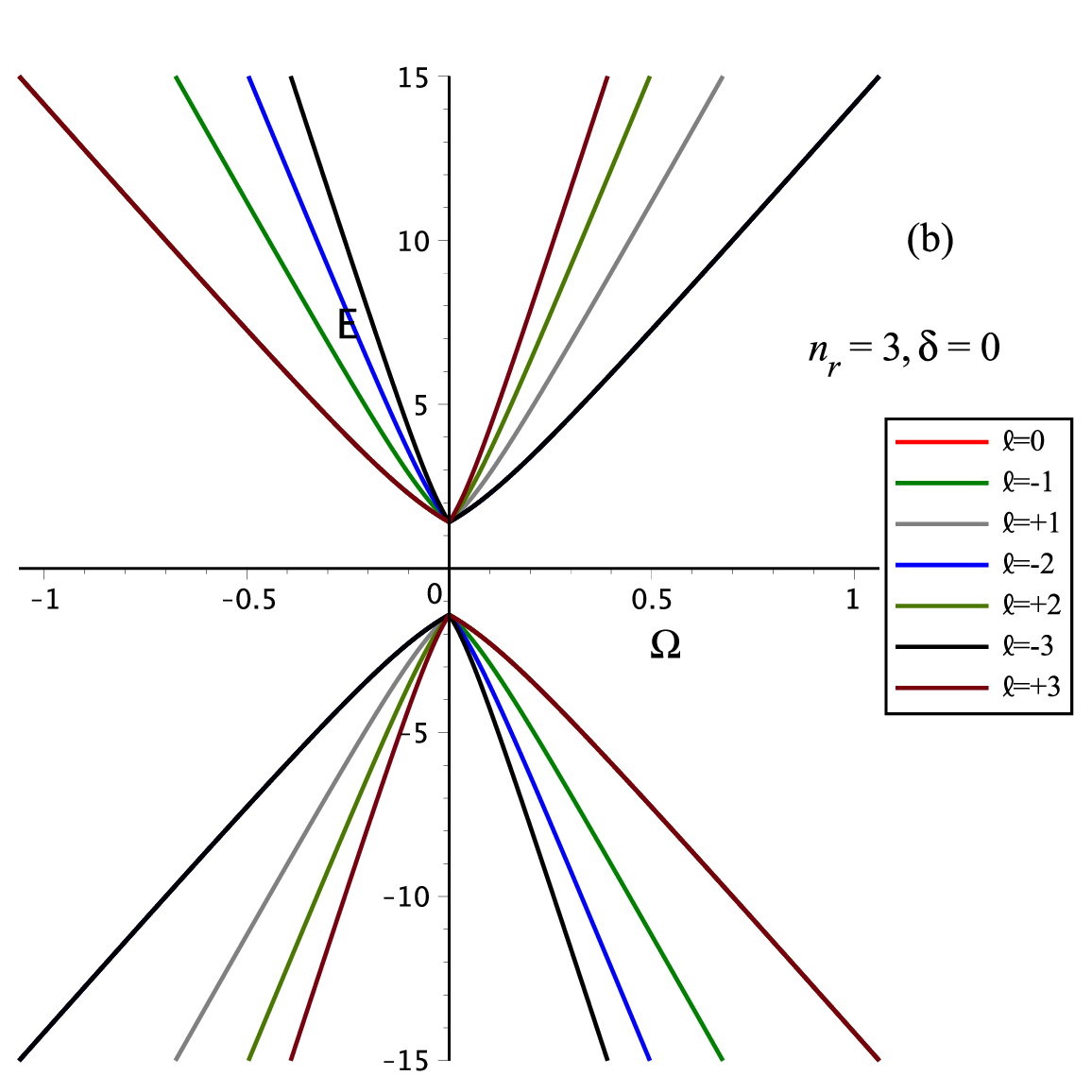} 
\includegraphics[width=0.3\textwidth]{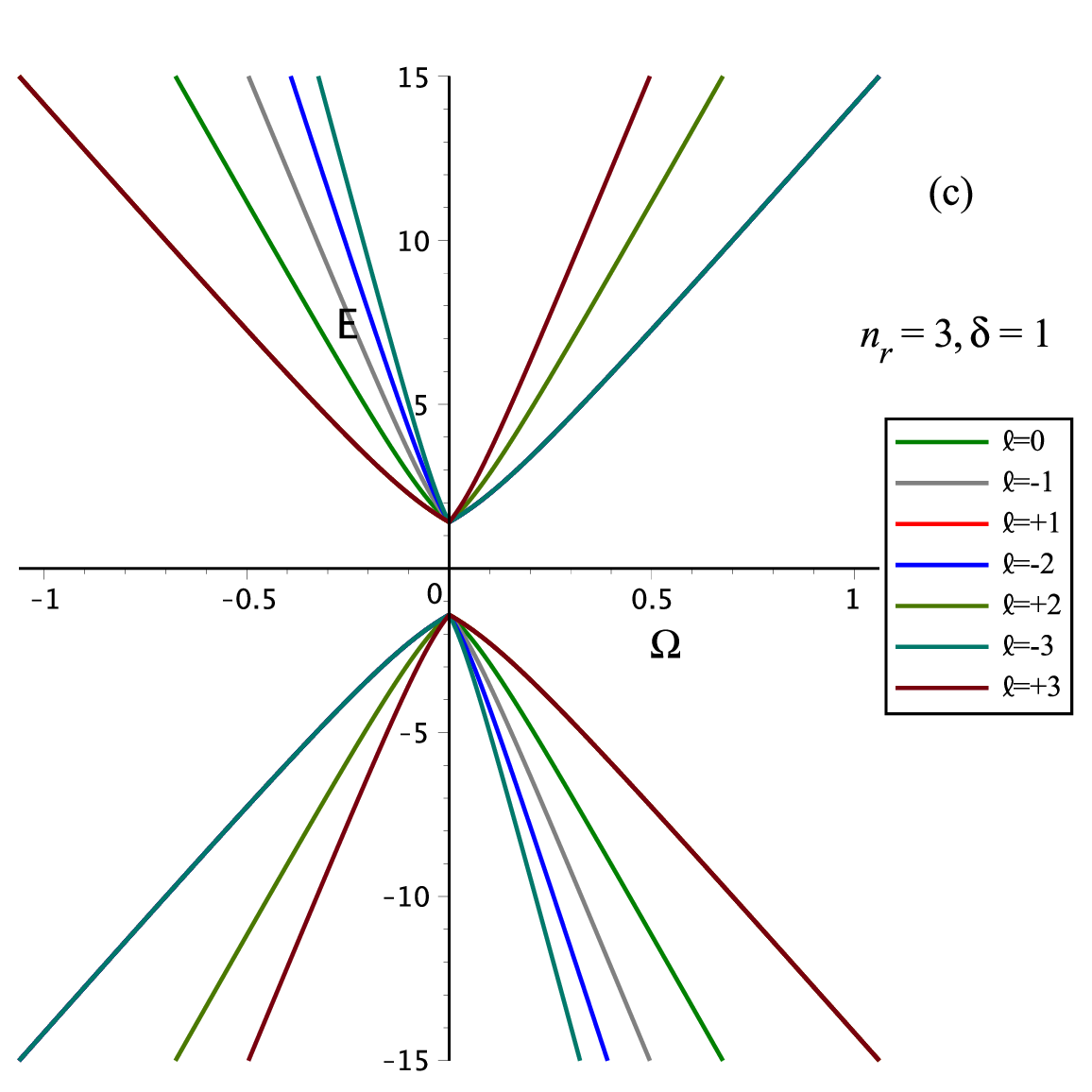}
\caption{\small 
{ The energy levels for KG-oscillators in SR-type spacetime of (\ref{e15.2}) are plotted with $\alpha=1/2, m_\circ=k=1, \tilde{\phi}=0$ (a) for $\delta=0$ (i.e., without space-like dislocation), $\ell=0$ and $n_{r}=0,1,2,3$, (b) for $\delta=0$, $n_{r}=3$ and $\ell=0,\pm1, \pm2,\pm3$, and (c) for $\delta=1$ (i.e., with space-like dislocation), $n_{r}=3$ and $\ell=0,\pm1, \pm2,\pm3$.}}
\label{fig1}
\end{figure}%
for $|\omega| =-\Omega _{\mp }E_{\pm }$. Whereas, the second way is to report the energies based on the vorticity $%
\Omega =\Omega _{\pm }=\pm \left\vert \Omega \right\vert $ so that the energies (\ref{e15.4}) and (\ref{e15.5}) yield%
\begin{equation}
E_{\pm ,\Omega _{+}}=\pm \left\vert \Omega \right\vert \,\mathcal{N}_{\pm
}\pm \sqrt{\Omega ^{2}\mathcal{N}_{\pm }^{2}+m_{\circ }^{2}+k^{2}},
\label{e17.2}
\end{equation}
for $\Omega=\Omega_+=+|\Omega|$, and%
\begin{equation}
E_{\pm ,\Omega _{-}}=\pm \left\vert \Omega \right\vert \,\mathcal{N}_{\mp
}\pm \sqrt{\Omega ^{2}\mathcal{N}_{\mp }^{2}+m_{\circ }^{2}+k^{2}}.
\label{e17.3}
\end{equation}
for $\Omega=\Omega_-=-|\Omega|$. This seems to be a more interesting way to express the related energy levels. Where, one may clearly observe vorticity-energy correlations between $E_{\pm ,\Omega _{+}}$ and $E_{\pm ,\Omega _{-}}$ so that, for a given radial quantum number $n_r$ we have the correlations 
\begin{equation}
\left.E_{+ ,\Omega _{+}}\right\vert_{\mathcal{L}=\mp \left\vert \mathcal{L}\right\vert} = \left.E_{+ ,\Omega_{-}}\right\vert_{ \mathcal{L}=\pm \left\vert \mathcal{L}\right\vert} \label{e17.4}
\end{equation}
and %
\begin{equation}
\left.E_{- ,\Omega _{+}}\right\vert_{\mathcal{L}=\mp \left\vert \mathcal{L}\right\vert} = \left.E_{- ,\Omega_{-}}\right\vert_{ \mathcal{L}=\pm \left\vert \mathcal{L}\right\vert} \label{e17.5} 
\end{equation}
Such correlations, (\ref{e17.4}) and (\ref{e17.5}), may very well serve quantum entanglement between the KG-particles at hand. 

To provide better understanding of the reported figures, one should notice that the energies for $E_{+ ,\Omega _{+}}$ and $E_{- ,\Omega _{+}}$ are represented by the first and second quarters, respectively, whereas those for $E_{- ,\Omega _{-}}$ and $E_{+ ,\Omega _{-}}$ are represented by the third and fourth quarters, respectively (e.g., Figures 1). Yet, in each quarter (i.e., for each branch of the four energy branches),  we also observe that  $\mathcal{N}_{\pm}=2n_{r}+1$ for $\forall \mathcal{L=\mp}\left\vert \mathcal{L}\right\vert$. This would, in turn, make all states with a given $n_{r}$ merge into the corresponding $S$-state  (i.e., $\ell =0$ state) as shown in Figures 1(a), and 1(b) for $\delta=0$, and into the corresponding  $\ell=1$ state (i.e., $P$-state) as shown in Figure 1(c).  In general, for a given $n_r$ all states merge into the corresponding $(\ell=\delta k)$-state in each quarter of the energy branches. This should be attributed to the result in (\ref{e15.3}). This kind of degeneracies is called spacetime associated degeneracies (STAD).  Moreover, such STADs should not be confused with the vorticity-energy correlations in (\ref{e17.4}) and (\ref{e17.5}).  Such correlations are anticipated to add new insights of the effects of gravitational fields on the quantum mechanical properties of relativistic particles (here, KG-particles), that may flourish research interest in the study of different aspects related to such findings.

\section{PDM KG-oscillators in SR-type cosmic string spacetime with space-like dislocation}

In this section, we consider a set of PDM KG-oscillators in the SR-type cosmic string spacetime described by (\ref{e11}), (\ref{e12}), and (\ref{e13}). We report, in what follows, vorticity-energy and charge-energy correlations .

\subsection{PDM KG-oscillators and vorticity-energy correlations}

A power-law type positive valued scalar multiplier $g\left( r\right)
=A\,r^{\upsilon }\Rightarrow M\left( r\right) =\upsilon ^{2}/16r^{2}$, with $S\left( r\right) =0) $ and $B_{\circ }=0$, one obtains%
\begin{equation}
\left\{ \partial _{r}^{2}-\frac{\left( \mathcal{\tilde{L}}^{2}-1/4\right) }{%
r^{2}}-\omega ^{2}r^{2}+\tilde{\lambda}\right\} R\left( r\right)
=0;\;\,\omega ^{2}=\Omega ^{2}E^{2},  \label{e17.1}
\end{equation}%
where%
\begin{equation}
\tilde{\lambda}=E^{2}-eB_{\circ }\,\mathcal{L}-\left( k^{2}+m_{\circ
}^{2}\right) ,\mathcal{\tilde{L}}^{2}=\left( \,\tilde{\ell}-\tilde{\delta}%
\,k\right) ^{2}+\frac{\upsilon ^{2}}{16}.  \label{e18}
\end{equation}%
Under such settings, with $\mathcal{\tilde{L}}^{2}=\left( \,\tilde{\ell}-%
\tilde{\delta}\,k\right) ^{2}+\upsilon ^{2}/16$, one obtains%
\begin{equation}
\lambda =2\,\left\vert \omega \right\vert \left( 2n_{r}+\left\vert \mathcal{%
\tilde{L}}\right\vert +1\right) ,\;R\left( r\right) \sim r^{\left\vert 
\mathcal{\tilde{L}}\right\vert +1/2}\exp \left( -\frac{\left\vert \omega
\right\vert r^{2}}{2}\right) L_{n_{r}}^{\left\vert \mathcal{\tilde{L}}%
\right\vert }\left( \left\vert \omega \right\vert r^{2}\right) ,  \label{e19}
\end{equation}%
and consequently one obtains two sets of energies%
\begin{equation}
E_{\pm ,1}=\pm \left\vert \Omega \right\vert \mathcal{\grave{N}}_{+}\pm 
\sqrt{\Omega ^{2}\,\mathcal{\grave{N}}_{+}^{2}+m^{2}+k^{2}}\Rightarrow
\left\{ 
\begin{tabular}{l}
$E_{+,1}=+\left\vert \Omega \right\vert \mathcal{\grave{N}}_{+}+\sqrt{\Omega
^{2}\,\mathcal{\grave{N}}_{+}^{2}+m^{2}+k^{2}}\smallskip $ \\ 
$E_{-,1}=-\left\vert \Omega \right\vert \mathcal{\grave{N}}_{+}-\sqrt{\Omega
^{2}\,\mathcal{\grave{N}}_{+}^{2}+m^{2}+k^{2}}$%
\end{tabular}%
\right. ,  \label{e20}
\end{equation}%
for $|\omega| =+\Omega _{\pm }E_{\pm }$, and%
\begin{equation}
E_{\pm ,2}=\pm \left\vert \Omega \right\vert \mathcal{\grave{N}}_{-}\pm 
\sqrt{\Omega ^{2}\,\mathcal{\grave{N}}_{-}^{2}+m^{2}+k^{2}}\Rightarrow
\left\{ 
\begin{tabular}{l}
$E_{+,2}=+\left\vert \Omega \right\vert \mathcal{\grave{N}}_{-}+\sqrt{\Omega
^{2}\,\mathcal{\grave{N}}_{-}^{2}+m^{2}+k^{2}\smallskip }$ \\ 
$E_{-,2}=-\left\vert \Omega \right\vert \mathcal{\grave{N}}_{-}-\sqrt{\Omega
^{2}\,\mathcal{\grave{N}}_{-}^{2}+m^{2}+k^{2}}$%
\end{tabular}%
\right. ,  \label{e21}
\end{equation}%
for $|\omega| =-\Omega _{\mp }E_{\pm }$. Where,%
\begin{equation}
\mathcal{\grave{N}}_{\pm }=\left( 2n_{r}+\left\vert \mathcal{\tilde{L}}%
\right\vert \pm \mathcal{L}+1\right). \label{e19.1}
\end{equation}%
However, in terms of the vorticities we may rearrange such energies so that%
\begin{equation}
E_{\pm ,\Omega _{+}}=\pm \left\vert \Omega \right\vert \,\mathcal{\grave{N}}%
_{\pm }\pm \sqrt{\Omega ^{2}\mathcal{\grave{N}}_{\pm }^{2}+m_{\circ
}^{2}+k^{2}},  \label{e21.2}
\end{equation}%
for $\Omega=\Omega_+=+|\Omega|$ and%
\begin{equation}
E_{\pm ,\Omega _{-}}=\pm \left\vert \Omega \right\vert \,\mathcal{\grave{N}}%
_{\mp }\pm \sqrt{\Omega ^{2}\mathcal{\grave{N}}_{\mp }^{2}+m_{\circ
}^{2}+k^{2}},  \label{e21.3}
\end{equation}
for $\Omega=\Omega_-=-|\Omega|$. Again, one clearly observes the vorticity-energy correlations between $E_{\pm ,\Omega _{+}}$ and $E_{\pm ,\Omega _{-}}$ so that, $E_{+ ,\Omega _{+}}|_{\mathcal{L}%
=\mp \left\vert \mathcal{L}\right\vert} = E_{+ ,\Omega
_{-}}|_{\mathcal{L}=\pm \left\vert \mathcal{L}\right\vert}$ and  $E_{- ,\Omega _{+}}|_{\mathcal{L}%
=\mp \left\vert \mathcal{L}\right\vert} =E_{- ,\Omega
_{-}}|_{\mathcal{L}=\pm \left\vert \mathcal{L}\right\vert}$ for a given radial quantum number $n_{r}$. However, we notice that the STAD degeneracies observed in each branch of the four energy branches, for constant mass (in the preceding section),  are removed/lifted by the PDM settings. This is an obvious consequence of  of (\ref{e19.1}), where in the PDM case  $\mathcal{\grave{N}}_{\pm}\neq2n_{r}+1$ for $\forall \mathcal{L=\mp}\left\vert \mathcal{L}\right\vert$.

\subsection{PDM KG-oscillators and charge-energy correlations}

A charged PDM KG-particle moving in G\"{o}del SR-type cosmic string spacetime with $\Omega =0$ (i.e., a non-rotating G\"{o}del universe, where metric (\ref{e2}) reduces into a cosmic string spacetime metric) with space-like dislocation and magnetic and Aharonov-Bohm flux fields of (\ref{e10}). In this case and for a positive valued scalar multiplier $g\left( r\right) =\exp (2\xi
r^{2})\Rightarrow M\left( r\right) =\xi ^{2}r^{2}+2\xi ;$ $\xi \geq 0$, we get%
\begin{equation}
\left\{ \partial _{r}^{2}-\frac{\left( \mathcal{L}^{2}-1/4\right) }{r^{2}}-%
\tilde{\Omega}^{2}r^{2}+\tilde{\lambda}\right\} R\left( r\right) =0;\;\,%
\tilde{\Omega}=\sqrt{\left( \frac{eB_{\circ }}{2}\right) ^{2}+\xi ^{2}},
\label{e21.1}
\end{equation}%
\begin{equation}
\;\tilde{\lambda}=E^{2}-eB_{\circ }\mathcal{L}-\left( k^{2}+m_{\circ
}^{2}+2\xi \right) =2\left\vert \tilde{\Omega}\right\vert \left(
2n_{r}+\left\vert \mathcal{L}\right\vert +1\right) .   \label{e23}
\end{equation}%
This would result in%
\begin{equation}
\tilde{\lambda}=\,\left\vert e\right\vert B_{\circ }\sqrt{1+\frac{4\xi ^{2}%
}{e^{2}B_{\circ }^{2}}}\,\left( 2n_{r}+\left\vert \mathcal{L}\right\vert
+1\right) ,\;R\left( r\right) \sim r^{\left\vert \mathcal{L}\right\vert
+1/2}\exp \left( -\frac{\left\vert \tilde{\Omega}\right\vert r^{2}}{2}%
\right) L_{n_{r}}^{\left\vert \mathcal{L}\right\vert }\left( \left\vert 
\tilde{\Omega}\right\vert r^{2}\right) ,  \label{24}
\end{equation}%
and consequently one obtains%
\begin{equation}
E=\pm \sqrt{k^{2}+m_{\circ }^{2}+2\xi +eB_{\circ }\mathcal{L+}\,\left\vert
e\right\vert B_{\circ }\sqrt{1+\frac{4\xi ^{2}}{e^{2}B_{\circ }^{2}}}%
\,\left( 2n_{r}+\left\vert \mathcal{L}\right\vert +1\right) }.  \label{e25}
\end{equation}%
Notably, with%
\begin{equation}
\mathcal{\tilde{N}}_{\pm }=\sqrt{1+\frac{4\xi ^{2}}{e^{2}B_{\circ }^{2}}}%
\,\left( 2n_{r}+\left\vert \mathcal{L}\right\vert +1\right) \pm \mathcal{L}
\label{e26}
\end{equation}%
two sets of energies are obtained%
\begin{equation}
E_{\pm ,+ve}=\pm \sqrt{m^{2}+k^{2}+2\xi +\left\vert e\right\vert B_{\circ }\,%
\mathcal{\tilde{N}}_{+}},  \label{e27}
\end{equation}%
for positively ($+ve$) charged PDM KG-particle and%
\begin{equation}
E_{\pm ,-ve}=\pm \sqrt{m^{2}+k^{2}+2\xi +\left\vert e\right\vert B_{\circ }\,%
\mathcal{\tilde{N}}_{-}},  \label{e28}
\end{equation}
for negatively ($-ve$) charged PDM KG-particle. We clearly observe that yet another type of correlations is introduced here, to be called, hereinafter, the charge-energy correlations, %
\begin{equation*}
E_{\pm ,+ve}|_{\mathcal{L=\pm }\left\vert 
\mathcal{L}\right\vert} =E_{\pm ,-ve}|_{\mathcal{L=\mp }%
\left\vert \mathcal{L}\right\vert}. \label{e28.1}
\end{equation*}
Again we see that the PDM removes/lifts STAD degeneracies in each branch of the energy levels. However, such degeneracies would reappear when $\xi=0$, where in this case  $\mathcal{\tilde{N}}_{\pm}=2n_{r}+1$ for  $\forall \mathcal{L=\mp}\left\vert \mathcal{L}\right\vert $

\begin{figure}[!ht]  
\centering
\includegraphics[width=0.3\textwidth]{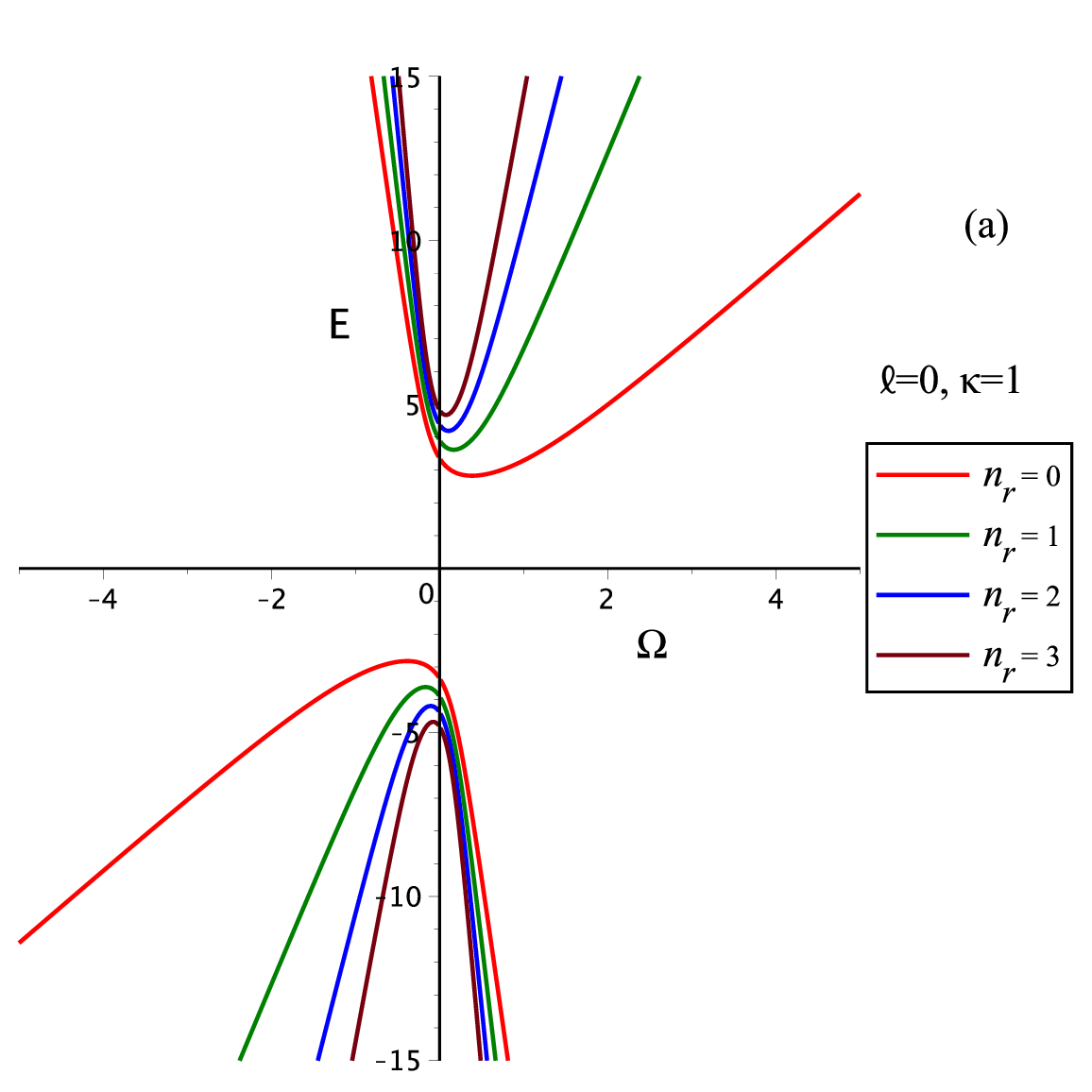}
\includegraphics[width=0.3\textwidth]{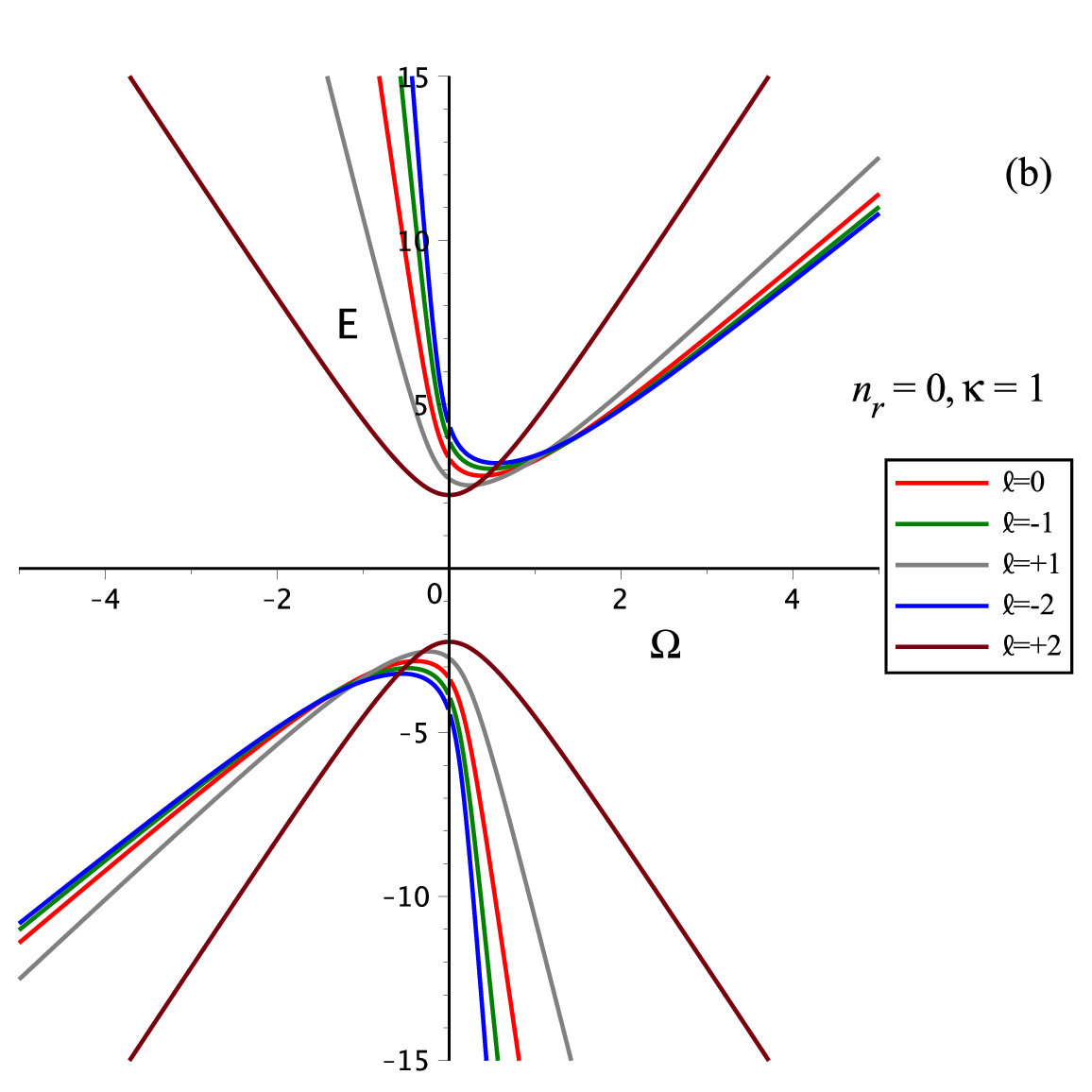} 
\includegraphics[width=0.3\textwidth]{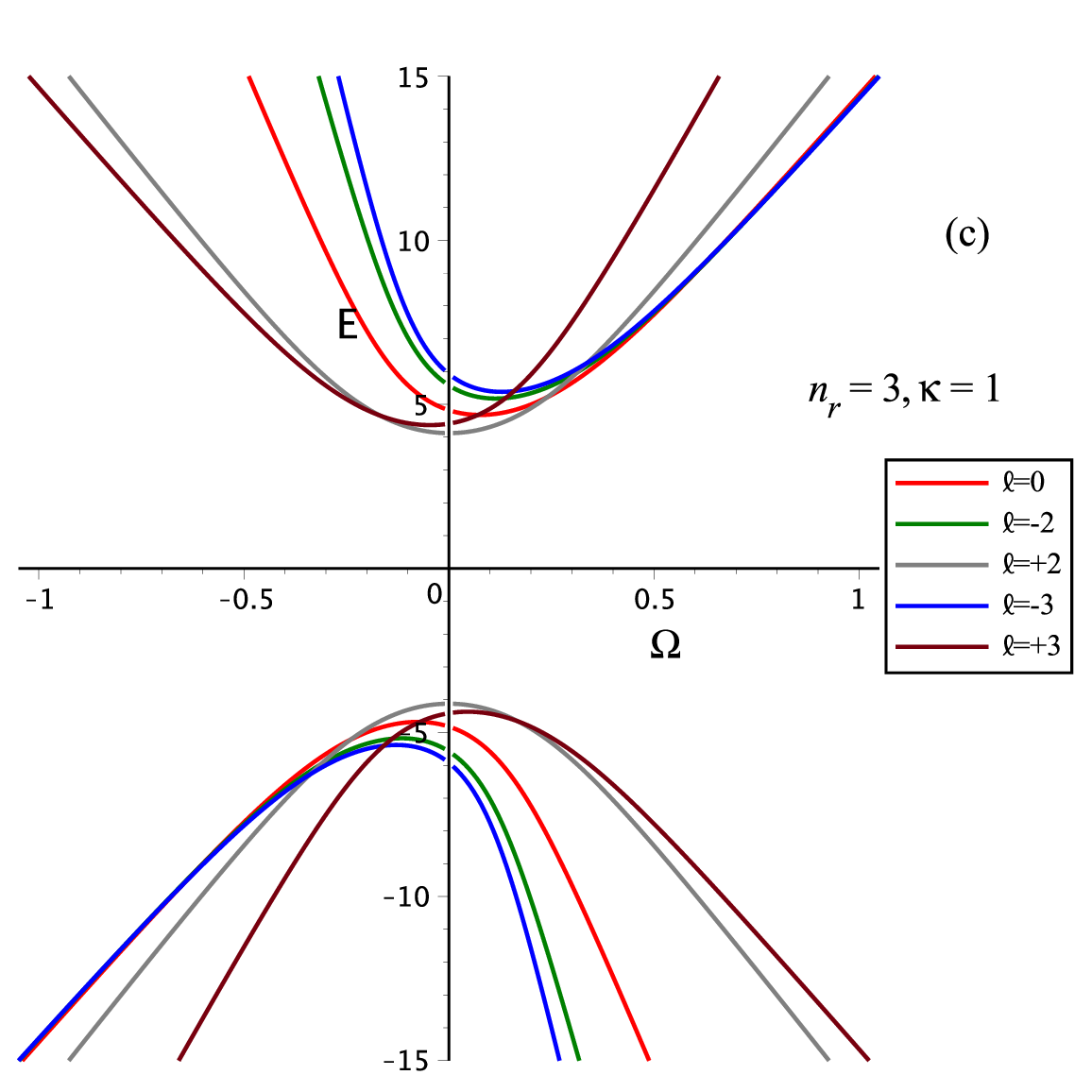}
\caption{\small 
{ The energy levels for PDM KG-oscillators in linear confinement  in SR-type spacetime with space-like dislocation of (\ref{e34}) are plotted with $\alpha=1/2, m_\circ=k=\tilde{\phi}=\delta=1$, and $\nu=4$ (a) for $\ell=0, \kappa=1$, and $n_{r}=0,1,2,3$, (b) for $\kappa=1$, $n_{r}=0$ and $\ell=0,\pm1, \pm2$, and (c) for $\kappa=1$, $n_{r}=3$ and $\ell=0,\pm2, \pm3$.}}
\label{fig2}
\end{figure}%

\section{PDM KG-oscillators in a linear confinement in SR-type cosmic string spacetime with space-like dislocation}

Lets us consider a charged PDM KG-particle in a SR-type cosmic string spacetime metric with space-like dislocation, in a linear Lorentz scalar potential $S\left( r\right) =\kappa\, r$ \cite{R39,R39.1}. We shall also assume that the positive-valued dimensionless scalar multiplier $g\left( r\right) =A\,r^{\upsilon }\Rightarrow
M\left( r\right) =\upsilon ^{2}/16r^{2}$, and $B_{\circ }=0$. In this case (\ref{e11}) reads%
\begin{equation}
\left\{ \partial _{r}^{2}-\frac{\left( \mathcal{\tilde{L}}^{2}-1/4\right) }{%
r^{2}}-\tilde{\omega}^{2}r^{2}-2m_{\circ }\,\kappa \,r+\grave{\lambda}%
\right\} R\left( r\right) =0,  \label{e29}
\end{equation}
where%
\begin{equation}
\grave{\lambda}=E^{2}-2\Omega E\mathcal{L}-\left( k^{2}+m_{\circ
}^{2}\right) ,\;\tilde{\omega}^{2}=\omega ^{2}+\kappa ^{2},\mathcal{L}%
^{2}=\left( \,\tilde{\ell}-\tilde{\delta}\,k\right) ^{2},\;\mathcal{\tilde{L}%
}^{2}=\mathcal{L}^{2}+\frac{\upsilon ^{2}}{16},  \label{e30}
\end{equation}

This is a two-dimensional radial Schr\"{o}dinger-like quantum model (\ref{e29}), where it admits a solution in the form of 
\begin{equation}
R\left( r\right) =r^{\left\vert 
\mathcal{\tilde{L}}\right\vert +1/2}\exp \left( -\frac{\left\vert \tilde{%
\omega}\right\vert r^{2}}{2}-\frac{\tilde{\kappa}}{2\,\left\vert \tilde{%
\omega}\right\vert }r\right) \,H_{B}\left( \alpha ^{\prime },\beta ^{\prime
},\gamma ^{\prime },\delta ^{\prime },\sqrt{\tilde{\omega}}r\right) ,
\label{e31}
\end{equation}
where $\,\tilde{\kappa}=2m_{\circ }\kappa $, 
\begin{equation}
\alpha ^{\prime }=2\left\vert \mathcal{\tilde{L}}\right\vert \,,\,\beta
^{\prime }=\frac{\tilde{\kappa}}{\left\vert \tilde{\omega}\right\vert ^{3/2}}%
\,,\,\gamma ^{\prime }=\frac{\grave{\lambda}}{\left\vert \tilde{\omega}%
\right\vert }+\frac{\tilde{\kappa}^2}{4\left\vert \tilde{\omega}\right\vert
^{3}}\,,\delta ^{\prime }=0,  \label{e32}
\end{equation}%
and $H_{B}\left( \alpha ^{\prime },\beta ^{\prime },\gamma ^{\prime },\delta
^{\prime },\sqrt{\tilde{\omega}}r\right) $ is the biconfluent Heun series. In order for the biconfluent Heun series to become a polynomial of degree $%
n\geq 0$, one has to apply the condition that $\gamma ^{\prime }=2\left(
n+1\right) +\alpha ^{\prime }$ \cite{R38,R39,R40,R41,R42,R43}. This condition would imply that%
\begin{equation}
\grave{\lambda}=2\left\vert \tilde{\omega}\right\vert \left( n+\left\vert 
\mathcal{\tilde{L}}\right\vert +1\right) -\frac{\tilde{\kappa}^{2}}{4\tilde{%
\omega}^{2}}=2\left\vert \tilde{\omega}\right\vert \left( 2n_{r}+\left\vert 
\mathcal{\tilde{L}}\right\vert +1\right) -\frac{\tilde{\kappa}^{2}}{4\tilde{%
\omega}^{2}},  \label{e33}
\end{equation}
where we have replaced the truncation order $n$ of the Heun series by $2n_{r};\,\,n_{r}=0,1,2,\cdots $, so that when $\kappa =0$ the eigenvalues of the pure oscillator of (\ref{e15}) are recovered. Therefore,  the truncation of the power series should be at $n=2n_{r}$, where $n_r$ is the radial quantum number \cite{R28,R29,R30} (to be discussed in more details in the Appendix section below). This result would allow us to obtain%
\begin{equation}
E^{2}-2\Omega E\mathcal{L}-\left( k^{2}+m_{\circ }^{2}\right) =2\left\vert
\Omega E\right\vert \sqrt{1+\frac{\kappa ^{2}}{\Omega ^{2}E^{2}}}\left(
2n_{r}+\left\vert \mathcal{L}\right\vert +1\right) -\frac{m_{\circ
}^{2}\kappa ^{2}}{\Omega ^{2}E^{2}+\kappa ^{2}}.  \label{e34}
\end{equation}%
\begin{figure}[!ht]  
\centering
\includegraphics[width=0.3\textwidth]{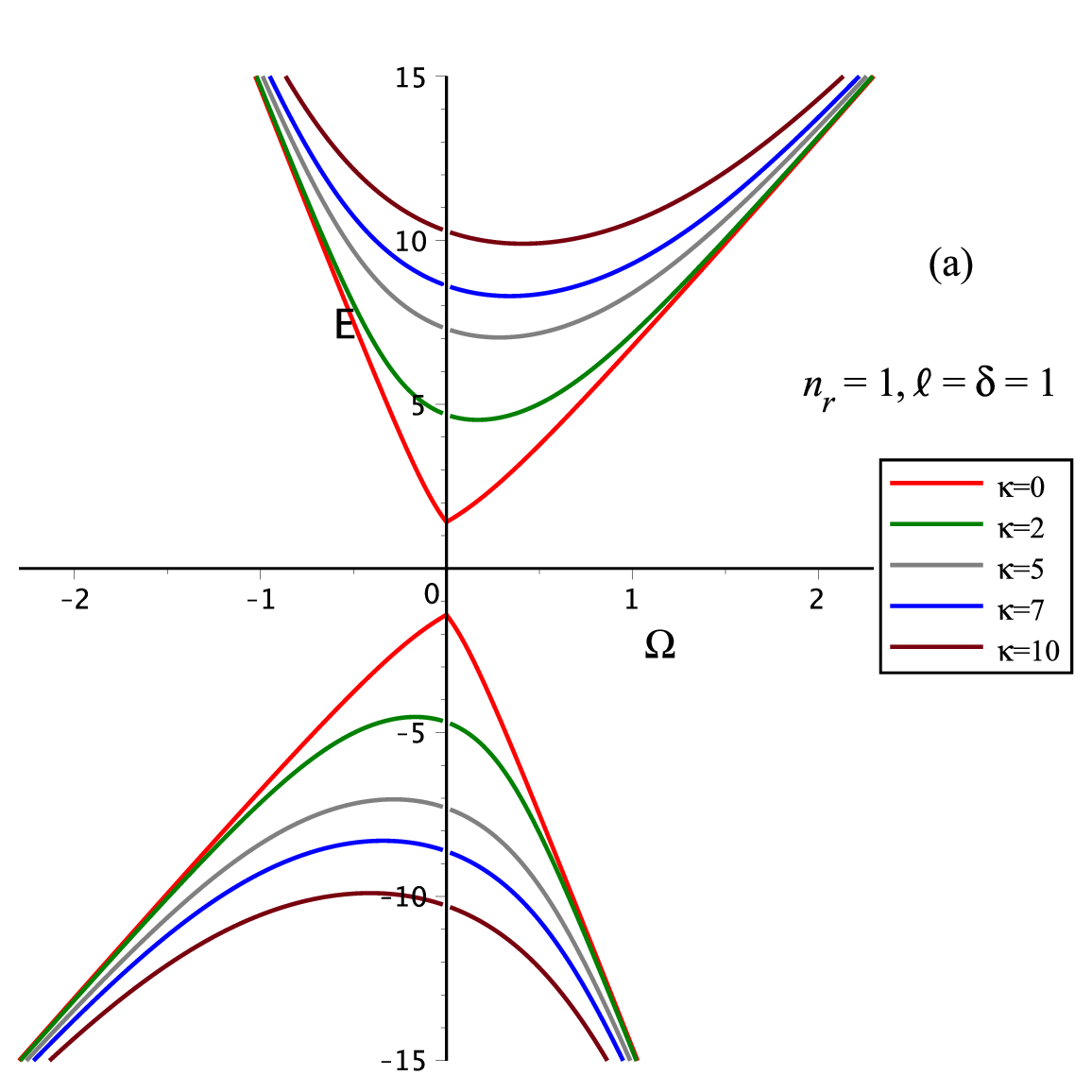}
\includegraphics[width=0.3\textwidth]{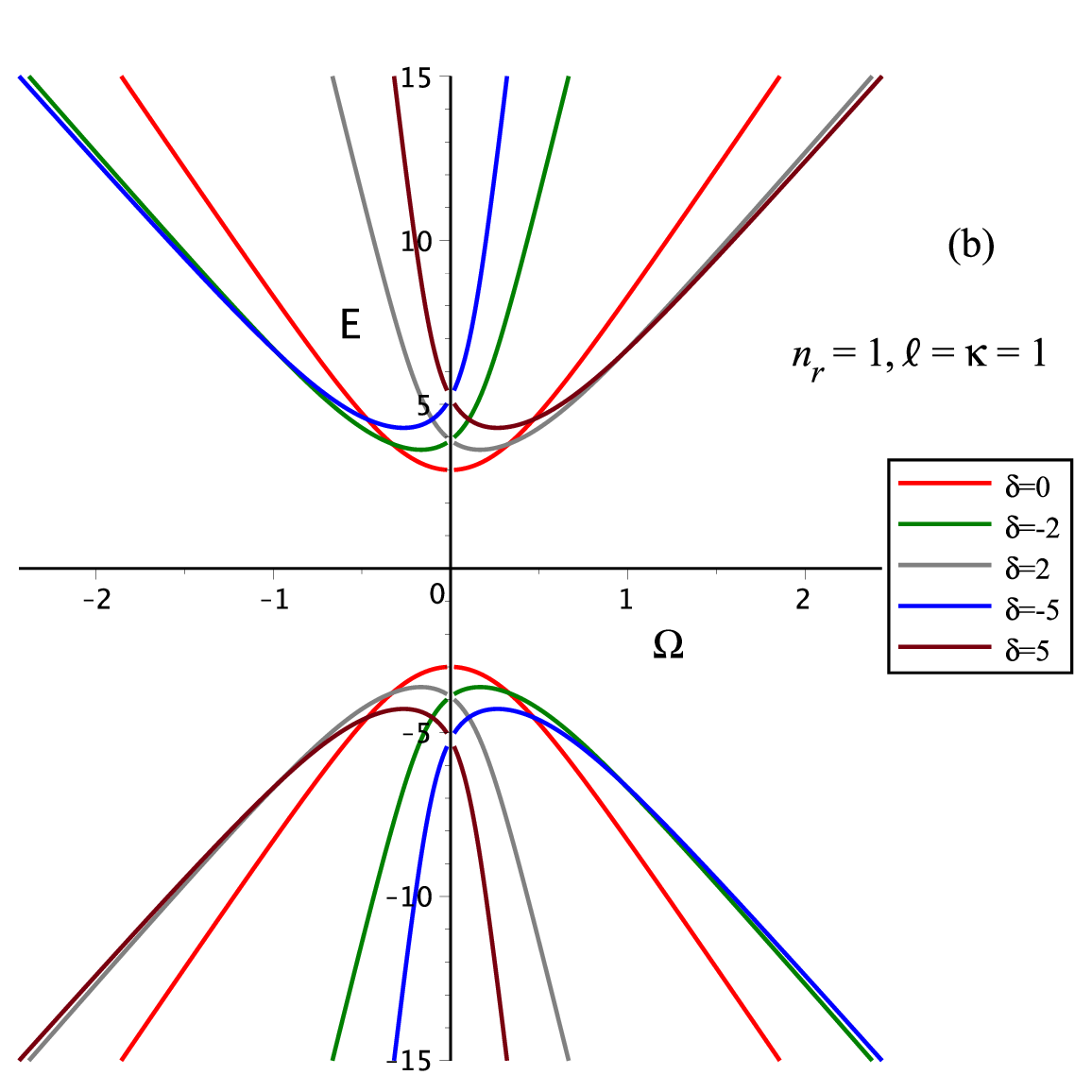} 
\includegraphics[width=0.3\textwidth]{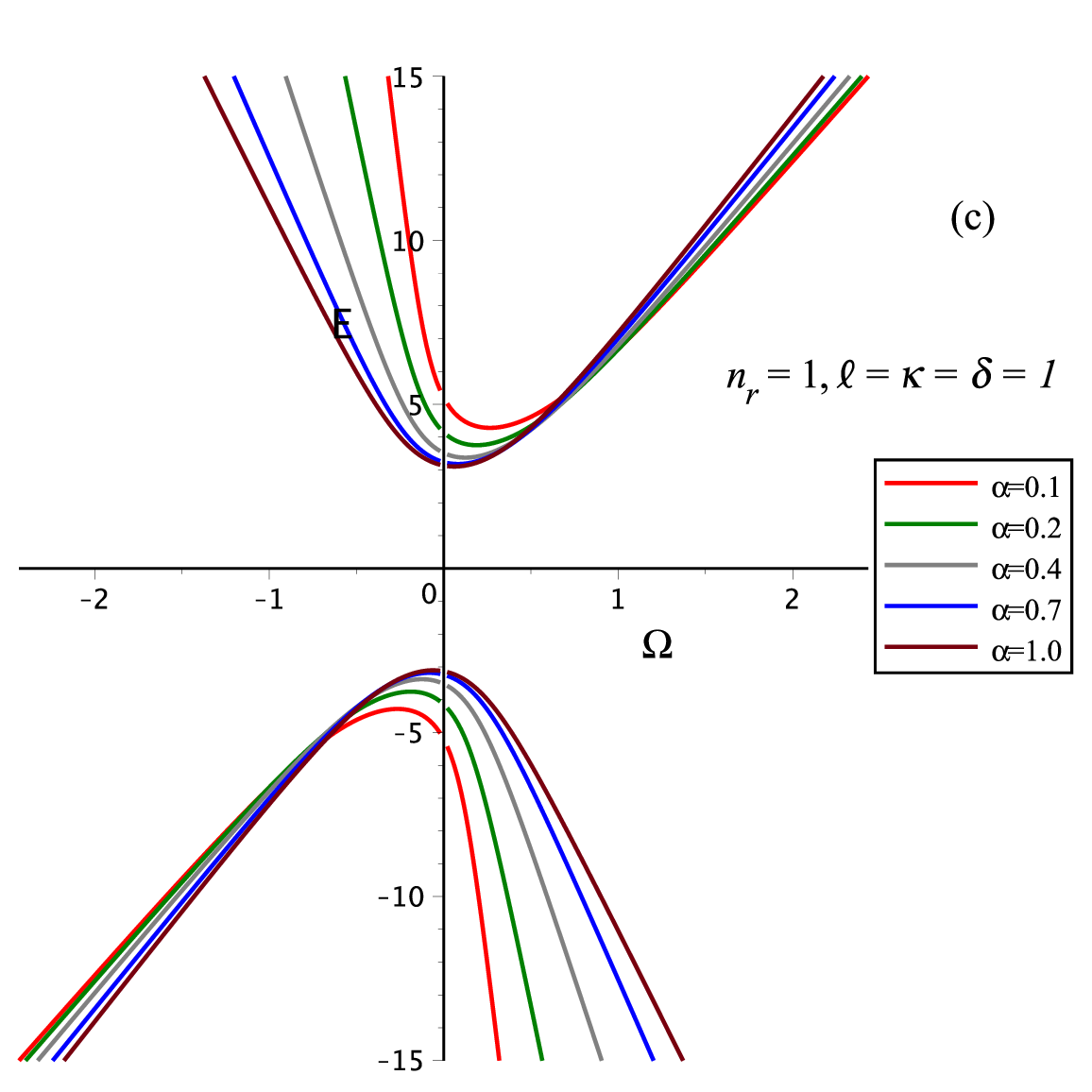}
\caption{\small 
{The energy levels for PDM KG-oscillators in linear confinement  in SR-type spacetime with space-like dislocation of (\ref{e34}) are plotted with $ m_\circ=k=\tilde{\phi}=1$, and $\nu=4$ (a) for $\alpha=1/2$, $n_{r}=1$, $\ell=1$, $\delta=1$, and $\kappa=0,2,5,7,10$, (b) for $\alpha=1/2$, $\kappa=1$, $n_{r}=1$, $\ell=1$, and $\delta=0,\pm2,\pm5$, and (c) for $\kappa=1=\delta$, $n_{r}=1$ and $\alpha=0.1, 0.2, 0.4, 0.7, 10$.}}
\label{fig3}
\end{figure}%
It is obvious that this result retrieves those in (\ref{e16}) and (\ref{e17}) when the linear confinement is switched off, i.e., $\kappa =0$. This is in fact the natural tendency of any physically admissible and viable solution of a more general problem. The corresponding energies of (\ref{e34}) are plotted in Figures 2 and 3. Where, the vorticity-energy correlations again read $E_{+ ,\Omega _{+}}|_{\mathcal{L}%
=\mp \left\vert \mathcal{L}\right\vert} = E_{+ ,\Omega
_{-}}|_{\mathcal{L}=\pm \left\vert \mathcal{L}\right\vert}$ and  $E_{- ,\Omega _{+}}|_{ \mathcal{L}=\mp \left\vert \mathcal{L}\right\vert} =E_{- ,\Omega
_{-}}|_{\mathcal{L}=\pm \left\vert \mathcal{L}\right\vert}$,  where $E_{\pm ,\Omega _{+}}$ are represented by the first and second quarters, whereas $E_{\pm
,\Omega _{-}}$ are represented by the third and fourth quarters of Figures 2 and 3. In the Appendix section below. we thoroughly report and discuss an exact solution of the model in  (\ref{e29})  for $\mathcal{\tilde{L}}=1/2$.

\section{Concluding remarks}
In this work, we have revisited KG-particles in SR-type cosmic string spacetime with space-like dislocation. We have discussed the gravitational field effects on the corresponding spectroscopic structure through the assumption that the vorticity of the SR-type spacetime may take two possible values, i.e.,  $\Omega=\pm|\Omega|$. This assumption has allowed us to express the energy levels in terms of the vorticity. Consequently, we have introduced/reported the so called vorticity-energy correlations in (\ref{e17.4})  and  (\ref{e17.5})  as new and interesting characterizations of the effect of the gravitational field. We have obtained four branches of energy levels that are introduced by the condition $|\Omega E|\geq0$  of (\ref{e15.2}), that is $|\Omega E|=\Omega_{\pm} E_{\pm}$ or $|\Omega E|=-\Omega_{\mp} E_{\pm}$. Yet, we have observed that in each branch of the energy levels, the irrational quantum number $\mathcal{N}_{\pm}$ of (\ref{e15.3}) manifestly suggests that  $\mathcal{N}_{+}=2n_{r}+1$ for $\forall \mathcal{L=-}%
\left\vert \mathcal{L}\right\vert $ and $\mathcal{N}_{-}=2n_{r}+1$ for $%
\forall \mathcal{L=+}\left\vert \mathcal{L}\right\vert $. This have made all states with a given $n_{r}$ merge with the corresponding $S$-state  (i.e., $\ell =0$ state) as shown in each quarter of the energies in Figures 1(a), 1(b), and 1(c).  We have called this kind of degeneracies as spacetime associated degeneracies (STAD).  Based on such findings above, we have discussed PDM KG-oscillators and reported their vorticity-energy correlations and their charge-energy correlations. We have observed that PDM settings have removed STADs degeneracies. Moreover, we have also considered PDM KG-oscillators in a linear confinement and reported, through Figures 2 and 3, the corresponding vorticity-energy correlations. To the best of our knowledge, such results and findings have never been discussed elsewhere. 

However, in connection with PDM KG-oscillators in a linear confinement in SR-type cosmic string spacetime with space-like dislocation, we have given this problem some particular attention and discussed it thoroughly in  the Appendix below. Therein, we have provided a brute-force-evidence that there is no additional quantization, related to the condition $C_{n+1}=0$ in the biconfluent Heun series truncation process, as claimed in \cite{R41,R46,R47}. 

Finally, such findings are anticipated to provide new insights on the effects of the gravitational fields (introduced by the spacetime structure at hand) on the quantum mechanical properties of relativistic particles in general, which in turn may flourish research interest in the study of different aspects in this direction.

\section{Appendix}
This section is intended to address the asymptotic behaviour of the radial wave function $R(r)$ as $r\rightarrow 0$ and provide a brute-force-evidence on the truncation order of the biconfluent Heun series into a polynomial of order $2n_r$ instead of $n$ in (\ref{e33}). 

\subsection{Exact solvability of Schr\"{o}dinger-like model (\ref{e29}) for $\mathcal{\tilde{L}}=1/2$}

We recall equation (\ref{e29}), with for $\mathcal{\tilde{L}}=1/2$, to obtain%
\begin{equation}
\left\{ \partial _{r}^{2}-\tilde{\omega}^{2}r^{2}-2m_{\circ }\,\kappa \,r+\grave{\lambda}%
\right\} R\left( r\right) =0.  \label{Ae1}
\end{equation}
One should be reminded that this equation is equivalent to the spherically symmetric radial Schr\"{o}dinger equation with the angular momentum quantum number $\ell=0$ in the central repulsive core $\ell(\ell+1)/r^2$. Let us use a radial wave function%
\begin{equation}
    R(r)=e^{-\frac{|\tilde{\omega}|\,r^2}{2}-\frac{m_{\circ }\,\kappa \,r}{|\tilde{\omega}|}}\left[C_1\,(r+\tilde{A})F(r)+C_2\,G(r)\right];\,\, \tilde{A}=\frac{A}{\tilde{\omega}^2}=\frac{m_{\circ }\,\kappa}{\tilde{\omega}^2}. \label{Ae2}
\end{equation}
Obviously, the asymptotic behaviour of $R(r)\rightarrow 0$ as $r\rightarrow \infty$ is taken into account through the exponential function in (\ref{Ae2}). Whereas, the asymptotic behaviour of $R(r)$ as $r\rightarrow 0$ is either  finite or zero, depending on the effective interaction potential at hand ( e.g., \cite{R48}). We shall instead let the general functions $F(r)$ and $G(r)$ have their say in the process. Hence, the substitution of $R(r)$, (\ref{Ae2}), in (\ref{Ae1}) implies that%
\begin{eqnarray}
    &&C_1 \,\left[(r+\tilde{A})\,\partial_{r}^2 -2|\tilde{\omega}|\left((r+\tilde{A})^2-\frac{1}{|\tilde{\omega}|}\right)\,\partial_r+[\grave{\lambda}-3|\tilde{\omega}|+\frac{A^2}{\tilde{\omega}^2}] \right]F(r) \\
    &&+\, C_2\,\left[(r+\tilde{A})\,\partial_{r}^2 -2|\tilde{\omega}|((r+\tilde{A})\,\partial_r+[\grave{\lambda}-|\tilde{\omega}|+\frac{A^2}{\tilde{\omega}^2}] \right]G(r)=0. \label{Ae3}
\end{eqnarray}
This would immediately suggest  that%
\begin{equation}
    \left[(r+\tilde{A})\,\partial_{r}^2 -2|\tilde{\omega}|\left((r+\tilde{A})^2-\frac{1}{|\tilde{\omega}|}\right)\,\partial_r+[\grave{\lambda}-3|\tilde{\omega}|+\frac{A^2}{\tilde{\omega}^2}] \right]F(r)=0, \label{Ae4}
\end{equation}
and%
\begin{equation}
    \left[(r+\tilde{A})\,\partial_{r}^2 -2|\tilde{\omega}|((r+\tilde{A})\,\partial_r+[\grave{\lambda}-|\tilde{\omega}|+\frac{A^2}{\tilde{\omega}^2}] \right]G(r)=0. \label{Ae5}
\end{equation}
Of course, to come out with the same result, one may use one solution at a time and then the general radial wave function would be the linear combination of the two as in (\ref{Ae2}).  Nevertheless, these two equations (\ref{Ae4}) and (\ref{Ae5}) can be transformed, using a change of variable of the form of $\eta=|\tilde{\omega}|(r+\tilde{A})^2$, to yield%
\begin{equation}
    \left[\eta \,\partial_{\eta}^2 +(\frac{3}{2}-\eta)\,\partial_{\eta}+\frac{1}{4|\tilde{\omega}|}[\grave{\lambda}-3|\tilde{\omega}|+\frac{A^2}{\tilde{\omega}^2}] \right]\, F(\eta)=0, \label{Ae6}
\end{equation}
and%
\begin{equation}
     \left[\eta \,\partial_{\eta}^2 +(\frac{1}{2}-\eta)\,\partial_{\eta}+\frac{1}{4|\tilde{\omega}|}[\grave{\lambda}-|\tilde{\omega}|+\frac{A^2}{\tilde{\omega}^2}] \right]\, G(\eta)=0. \label{Ae7}
\end{equation}
Obviously, both equations are in the form of Kummer's equation%
\begin{equation}
    [\eta\,\partial_{\eta}^2+(b-\eta)\,\partial_\eta-a]W(\eta)=0.
\end{equation}
This would suggest that%
\begin{equation}
    F(\eta)=\,_{1}F_{1}(-\frac{1}{4|\tilde{\omega}|}(\grave{\lambda}-3|\tilde{\omega}|+\frac{A^2}{\tilde{\omega}^2}),\frac{3}{2},\eta), \label{Ae8}
\end{equation}
and%
\begin{equation}
     G(\eta)=\,_{1}F_{1}(-\frac{1}{4|\tilde{\omega}|}(\grave{\lambda}-|\tilde{\omega}|+\frac{A^2}{\tilde{\omega}^2}),\frac{1}{2},\eta), \label{Ae9}
\end{equation}
In this case, our $R(r)$ in (\ref{Ae2}) takes it most general form%
\begin{eqnarray}
R(r)=e^{-\frac{|\tilde{\omega}|\,r^{2}}{2}-\frac{m_{\circ }\,\kappa \,r}{|%
\tilde{\omega}|}}&& \left[C_{1}(r+\tilde{A})\,_{1}F_{1}(-\frac{1}{4|\tilde{\omega}|}(\grave{%
\lambda}-3|\tilde{\omega}|+\frac{A^{2}}{\tilde{\omega}^{2}}),\frac{3}{2}%
,\eta )\right.   \nonumber \\
&&\left. +\,C_{2}\,\,_{1}F_{1}(-\frac{1}{4|\tilde{\omega}|}(\grave{\lambda}-|%
\tilde{\omega}|+\frac{A^{2}}{\tilde{\omega}^{2}}),\frac{1}{2},\eta )\right].
\label{Ae10}
\end{eqnarray}
Now, we need to appeal to the exact textbook radial solution for the pure harmonic oscillator to figure out the exact form of our solution in (\ref{Ae10}). To do that, we take $\kappa=0\rightarrow \tilde{A}=0$ and require that $R(r)\rightarrow 0$ as $r\rightarrow 0$ (as textbook documented  behaviour). It is obvious that we have to take $C_{2}=0$. In this case, our exact solution to the problem in (\ref{Ae1}) is%
\begin{equation}
    R(r)=C\,(r+\tilde{A})\,e^{-\frac{|\tilde{\omega}|\,r^{2}}{2}-\frac{m_{\circ }\,\kappa \,r}{|%
\tilde{\omega}|}} \,_{1}F_{1} \left(-\frac{1}{4|\tilde{\omega}|}(\grave{%
\lambda}-3|\tilde{\omega}|+\frac{A^{2}}{\tilde{\omega}^{2}}),\frac{3}{2}%
,\, |\tilde{\omega}|(r+\tilde{A})^2 \right).
\label{Ae11}
\end{equation}
This solution suggests, beyond doubt, that a quantum particle moving in an oscillator-plus-linear radial potential is allowed to reach the center at $r=0$ with some amplitude ( e.g., \cite{R48}). Then the conjecture  that $R(r)\rightarrow 0$ as $r\rightarrow 0$ is not a general admissible condition in quantum mechanics, but finiteness and square integrability of the radial wave function is the only valid condition in general. 

Next, we need to terminate the confluent hypergeometric series of (\ref{Ae11}) into a polynomial of degree $n_r=0,1,2,\cdots$ by the requirement that%
\begin{equation}
-n_r=  -\frac{1}{4|\tilde{\omega}|}(\grave{\lambda}-3|\tilde{\omega}|+\frac{A^{2}}{\tilde{\omega}^{2}})\Rightarrow \grave{\lambda}=2|\tilde{\omega}|(2n_r+\frac{3}{2})-\frac{A^{2}}{\tilde{\omega}^{2}}. \label{Ae12}
\end{equation}
This is exactly the same as the result we have reported in (\ref{e33}) for $\mathcal{\tilde{L}}=1/2$. 

We now recollect our reported radial wave function of (\ref{e31}) and compare it with the one in (\ref{Ae11}) for $\mathcal{\tilde{L}}=1/2$. Such comparison would yield (up to a constant value) an interesting relation between the biconfluent Heun and the confluent hypergeometric functions/polynomials. Namely, %
\begin{equation}
(y+\frac{\beta^{\prime}}{2})\, _{1}F_{1} \left( \frac{3-\gamma^{\prime}}{4},\, \frac{3}{2},(y+\frac{\beta^{\prime}}{2})^2 \right) =  y\, H_{B}(1, \beta^{\prime}, \gamma^{\prime},0,y);\,\, y=\sqrt{|\tilde{\omega}|}\,r, \label{Ae13}
\end{equation}
where $\gamma^{\prime}$ and $\beta^{\prime}$ are given in  (\ref{e32}).  Whilst, the condition for a biconfluent Heun polynomial (that $\gamma ^{\prime }=2\left(n+1\right) +\alpha ^{\prime }$ , where $\alpha ^{\prime}=1$ in this case) would result that $\gamma ^{\prime }=2n+3$, the condition for a confluent hypergeometric polynomial gives  $\gamma ^{\prime }=4n_r+3$.  This immediately suggests that the truncation order for the biconfluent Heun polynomial is at $n=2n_r$ to yield our result in (\ref{e33}) and in (\ref{Ae12}). Obviously, both the confluent hypergeometric and the biconfluent Heun polynomials suggest that they are both of even powers.  This would justify our assumption for the truncation order in (\ref{e33})  and yet it presents, beyond doubt, an exact solution for the Schr\"{o}dinger problem in (\ref{Ae1}) (i.e., a special case for (\ref{e29})). Yet, our result in (\ref{Ae13}) is consistent with that from the well know relation, \cite{R43},%
\begin{equation}
H_B(\alpha ^{\prime },0,\gamma ^{\prime },0,y)=\,_{1}F_{1}\left(\frac{1}{2}+\frac{\alpha ^{\prime }}{4}-\frac{\gamma ^{\prime }}{4}, 1+\frac{\alpha ^{\prime }}{2},y^2\right). \label{Ae14}
\end{equation}
To observe this consistency, one would set $\beta^{\prime}=0$ in (\ref{Ae13}) and obtain%
\begin{equation}
 H_{B}(1, 0, \gamma^{\prime},0,y) = \, _{1}F_{1} \left( \frac{3-\gamma^{\prime}}{4},\, \frac{3}{2},y^2 \right)  . \label{Ae15}  
\end{equation}
Therefore, one would anticipate that there is an eminent and intimate relation between the biconfluent Heun and  the confluent hypergeometric functions/polynomials which is yet-to-be discovered. More interestingly, our result (\ref{Ae13}) and the relations  (\ref{Ae14}) and (\ref{Ae15}) provide a concrete evidence that biconfluent Heun polynomial are of even powers and the corresponding power series should be truncated into a polynomial of order $2n_r$ (instead of $n\geq 0$) using the condition that $\gamma ^{\prime }=2\left(2n_r+1\right) +\alpha ^{\prime }$.

The above analysis show, beyond doubt, that the radial wave function, of (\ref{Ae11}), has a finite amplitude at $r=0$, which is indeed an acceptable solution. The finiteness and square integrability of the radial wave function are the only valid conditions to be used here.  Hence, it is not true that the radial wave function should always vanish at $r=0$ for the harmonic oscillator plus a liner potentials. Should a pure radial oscillator be the problem at hand, then $R(r)\rightarrow 0$ as $r\rightarrow 0$. But the current problem is for the interaction potential $V_{eff.}(r)=\tilde{\omega}^2r^2+2Ar=\tilde{\omega}^2(r+A/\tilde{\omega}^2)^2-A^2/\tilde{\omega}^2$, which is effectively a shifted-by-a-constant harmonic oscillator.  This would, in turn, affect the asymptotic behaviour of the radial wave function as $r\rightarrow 0$.  Of course when $A=0$, $R(r)\rightarrow 0$ as $r\rightarrow 0$. However, for $A\neq 0$, one would follow a change of variable $\rho=r+A/\tilde{\omega}^2$ so that the effective interaction potential becomes $V_{eff.}(\rho)=\tilde{\omega}^2\rho^2-A^2/\tilde{\omega}^2$. Then, instead of  $r^{\left\vert \mathcal{\tilde{L}}\right\vert+1/2}$ we shall use $\rho^{\left\vert \mathcal{\tilde{L}}\right\vert+1/2}$ to describe the asymptotic behaviour of the redial wave function as $\rho\rightarrow A/\tilde{\omega}^2$ (i.e., $r\rightarrow 0$ as documented in (\ref{Ae11}) above). One should also keep in mind that the central repulsive core $\ell(\ell+1)/r^2\rightarrow \ell(\ell+1)/(\rho -A/\tilde{\omega}^2)^2$ and hence the singularity is no more at $r=0$ of the spherically symmetric Schr"{o}dinger oscillator but rather at $\rho= A/\tilde{\omega}^2$.

\subsection{Validity and/or admissibility of the biconfluent Heun series/polynomials conditions}

At this point, it is unavoidably inviting to question the quantum mechanical validity and/or admissibility of the biconfluent Heun series/polynomials conditions for the PDM KG-oscillators in a linear confinement (i.e., the Schr\"{o}dinger-like model in (\ref{e29})). In so doing, we substitute%
\begin{equation}
R\left( r\right) =N\,r^{\left\vert \mathcal{\tilde{L}}\right\vert
+1/2}\exp \left( -\frac{\left\vert \tilde{\omega}\right\vert r^{2}}{2}-\frac{%
\tilde{\kappa}}{2\,\left\vert \tilde{\omega}\right\vert }r\right) \,H\left( 
\sqrt{\,\left\vert \tilde{\omega}\right\vert }r\right) ;\;H\left( \sqrt{%
\,\left\vert \tilde{\omega}\right\vert }r\right) =H\left( \rho \right)
=\sum_{j=0}^{\infty }A_{j}\rho ^{j},  \label{e35}
\end{equation}
in (\ref{e29}) to obtain $A_{1}=\tilde{\kappa}/2\,\left\vert \tilde{\omega}%
\right\vert ^{3/2}$, $A_{0}=1$, and%
\begin{equation}
A_{j+2}\,\left( j+2\right) \left( j+2\left\vert \mathcal{\tilde{L}}%
\right\vert +2\right) =A_{j+1}\left[ A_{1}\left( 2j+2\left\vert \mathcal{%
\tilde{L}}\right\vert +3\right) \right] +A_{j}\left( 2\,j-\frac{\tilde{%
\lambda}}{\left\vert \tilde{\omega}\,\right\vert }\right) ;\;j=0,1,2,\cdots, 
\label{e36}
\end{equation}%
where 
\begin{equation}
\tilde{\lambda}=\grave{\lambda}+\frac{\tilde{\kappa}^{2}}{4\,\tilde{\omega}%
^{2}}-2\,\left\vert \tilde{\omega}\right\vert \left( 1+\left\vert \mathcal{%
\tilde{L}}\right\vert \right) .  \label{e37}
\end{equation}%
Next, the biconfluent Heun series should be terminated at some $j=n$ to become a polynomial of degree $n\geq 0$ to secure finiteness and square integrability of the corresponding radial wavefunctions. Then, one would require that for $\forall j>n$ we have $A_{n+1}=0=A_{n+2}$ to imply 
\begin{equation}
A_{n}\left( 2n-\frac{\tilde{\lambda}}{\,\left\vert \tilde{\omega}%
\,\right\vert }\right) =0\Longrightarrow \tilde{\lambda}=2\,\left\vert 
\tilde{\omega}\,\right\vert n\Longrightarrow \grave{\lambda}=2\,\left\vert 
\tilde{\omega}\right\vert \left( n+\left\vert \mathcal{\tilde{L}}\right\vert
+1\right) -\frac{m_{\circ }^{2}\kappa ^{2}}{\tilde{\omega}^{2}};\;n\geq 0.
\label{e38}
\end{equation}%
(this condition is, in fact, the same condition used to obtain (46) which is in exact accord with (51)). This would allow us to cast our (\ref{e35}) as%
\begin{equation}
R\left( r\right) =N\,r^{\left\vert \mathcal{\tilde{L}}\right\vert
+1/2}\exp \left( -\frac{\left\vert \tilde{\omega}\right\vert r^{2}}{2}-\frac{%
\tilde{\kappa}}{2\,\left\vert \tilde{\omega}\right\vert }r\right)
\,\sum_{j=0}^{n}A_{j}\left[ \sqrt{\,\left\vert \tilde{\omega}\right\vert }%
r\right] ^{j}.  \label{e38-1}
\end{equation}

Nevertheless, it has become a traditional practice for some authors to claim and/or believe that the condition $A_{n+1}=0$ should be used again to find out additional $n$-dependent quantization recipes or correlations between the effective potential parameters  (e.g., \cite{R41,R42,R43} and references cited therein).  Such practices have resulted in the what should be called, hereinafter, "quantum mechanical mess" (see the details below). Hereby, the three terms recursion relation (\ref{e36}) is rewritten (following e.g., \cite{R41,R42,R43}) as%
\begin{equation}
A_{j+1}\,\left( j+1\right) \left( j+2\left\vert \mathcal{\tilde{L}}%
\right\vert +1\right) =A_{j}\left[ A_{1}\left( 2j+2\left\vert \mathcal{%
\tilde{L}}\right\vert +1\right) \right] +A_{j-1}\left( 2\,\left\vert \tilde{%
\omega}\,\right\vert \left[ j-1-n\right] \right) ;\,\,A_{-1}=0,  \label{e39}
\end{equation}%
so that for $\forall j>n$ we have $A_{n+1}=0$ to imply that%
\begin{equation}
A_{n}\left[ A_{1}\left( 2\,n+2\left\vert \mathcal{\tilde{L}}\right\vert
+1\right) \right] =2\,\left\vert \tilde{\omega}\,\right\vert A_{n-1};\;n\geq
0,\,A_{-1}=0.  \label{e40}
\end{equation}%
Which, for $n=0$, $A_{0}=1$ and $A_{1}=\tilde{\kappa}/2\,\left\vert \tilde{%
\omega}\right\vert ^{3/2}\neq 0$, gives 
\begin{equation}
A_{0}\left( 1+2\left\vert \mathcal{\tilde{L}}\right\vert \right)
=2\,\left\vert \tilde{\omega}\,\right\vert A_{-1}=0\Longrightarrow
1+2\left\vert \mathcal{\tilde{L}}\right\vert =0\Longrightarrow \left\vert 
\mathcal{\tilde{L}}\right\vert =-\frac{1}{2}.  \label{e41}
\end{equation}%
This result is neither mathematically nor physically an acceptable one. Although straightforward, there is no need to continue for $n=1,2,\cdots $, therefore. Of course, it could be mathematically interesting to know that the coefficients of $A_{n+1}=0$ are obvious polynomials of degree $n+1$ in $A_{1}=\tilde{\kappa}%
/2\,\left\vert \tilde{\omega}\right\vert ^{3/2}$ (e.g., \cite{R41,R42,R43,R46,R47} ). However, one should be aware that $A_{n+1}$ is no longer involved in our solution (\ref{e38-1}), and hence whatever it entails does not belong to our acceptable quantum mechanical solution. 

All arguments given in the Appendix above should lead us to one conclusion. As long as quantum mechanics is in point,  the repeated usage of the condition  $A_{n+1}=0$  should be labeled, hereinafter, as \emph{a quantum mechanically redundant condition}. This would render the biconfluent Heun series/polynomials approach quantum mechanically useful and reliable. This is, in fact, what the biconfluent Heun series/polynomials approach can, quantum mechanically, safely offer, at its best.

\end{document}